\documentclass{emulateapj}
\slugcomment{Accepted by {\it The Astrophysical Journal}}
\def\fdg{\hbox{$.\!\!^\circ$}}

\def\farcs{\hbox{$.\mkern-4mu^{\prime\prime}$}}

\def\la{\mathrel{\hbox{\rlap{\hbox{\lower4pt\hbox{$\sim$}}}\hbox{$<$}}}}
\def\ga{\mathrel{\hbox{\rlap{\hbox{\lower4pt\hbox{$\sim$}}}\hbox{$>$}}}}

\shortauthors{Park}
\shorttitle{SGR0526--66}

\begin{document}
%\title{Spectral Nature of Quiescent X-Ray Emission from SGR 0526--66 in the Large 
%Magellanic Cloud}
\title{{\it NuSTAR} Detection of Quiescent Hard X-Ray Emission from SGR 0526--66 
in the Large Magellanic Cloud}

\author{Sangwook Park\altaffilmark{1,4}, Jayant Bhalerao\altaffilmark{1},
Oleg Kargaltsev\altaffilmark{2}, and
Patrick O. Slane\altaffilmark{3}} 

\altaffiltext{1}{Box 19059, Department of Physics, University of Texas at Arlington,
Arlington, TX 76019}
\altaffiltext{2}{Department of Physics, George Washington University,
725 21st Street NW, Washington, DC 20052}
%\altaffiltext{2}{Department of Physics and Astronomy, Rutgers University,
%136 Frelinghuysen Road, Piscataway, NJ 08854-8019}
\altaffiltext{3}{Harvard-Smithsonian Center for Astrophysics, 60 Garden Street,
Cambridge, MA 02138}
\altaffiltext{4}{s.park@uta.edu}
%\altaffiltext{4}{Department of Astronomy and Astrophysics, Pennsylvania State
%University, 525 Davey Laboratory, University Park, PA 16802}
%\altaffiltext{5}{Korea Astronomy and Space Science Institute, Daejeon, 305-348, Korea}
%\altaffiltext{6}{Department of Applied Physics, University of Miyazaki, 1-1 Gakuen 
%Kibana-dai Nishi, Miyazaki, 889-2192, Japan} 

%\altaffiltext{1}{s.park@uta.edu}

\begin{abstract}

The soft $\gamma$-ray repeater (SGR) 0526--66 is the first-identified magnetar, 
and is projected within the supernova remnant N49 in the Large Magellanic Cloud. 
Based on our $\sim$50 ks {\it NuSTAR} observation, we detect the quiescent-state 
0526--66 for the first time in the 10--40 keV band. Based on the joint analysis 
of our {\it NuSTAR} and the archival {\it Chandra} ACIS data, we firmly establish 
the presence of the nonthermal component in the X-ray spectrum of 0526--66  
in addition to the thermal emission. In the best-fit blackbody (BB) plus power law
(PL) model, the slope of the PL component (photon index $\Gamma$ = 2.1) 
is steeper than those ($\Gamma$ $\la$ 1.5) for other magnetars. 
The soft part of the X-ray spectrum can be 
described with a BB component with the temperature of $kT$ = 0.43 keV. The 
best-fit radius ($R$ = 6.5 km) of the X-ray-emitting area is smaller than 
the canonical size of a neutron star. If we assume an underlying cool BB 
component with the canonical radius of $R$ = 10 km for the neutron star in 
addition to the hot BB component (2BB + PL model), a lower BB temperature 
of $kT$ = 0.24 keV is obtained for the passively cooling neutron star's
surface, while the hot spot emission with $kT$ = 0.46 keV dominates the 
thermal spectrum ($\sim$85\% of the thermal luminosity in the 0.5--5 keV 
band). The nonthermal component ($\Gamma$ $\sim$ 1.8) is still required.  

%Although it is speculative in the 
%current data, the softness of the broadband X-ray spectrum and the high 
%temperature for the thermal component (presumably emission from the 10$^{3-4}$ 
%yr-old neutron star's cooling) may suggest a strong effect from the high 
%magnetic fields on the cooling of this magnetar in the relatively long quiescent 
%period of $\sim$40 years.

\end{abstract}


\section {\label {sec:intro} INTRODUCTION}

The soft $\gamma$-ray repeater (SGR) 0526--66 showed intense $\gamma$-ray outbursts 
on 1979 March 5 and a luminous pulsed afterglow with $P$ $\sim$ 8 s \citep{maze79,clin80}, 
which led to the discovery of the magnetar phenomenon. The quiescent soft X-ray 
counterpart is projected within the supernova remnant (SNR) N49 in the Large Magellanic
Cloud (LMC) \citep{roth94}, and its pulsations with $P$ = 8.0436 s and $\dot{P}$
= 6.6 $\times$ 10$^{-11}$ s s$^{-1}$ were detected with 
a low pulsed-fraction of $f_p$ $\sim$ 10\% in the {\it Chandra} data \citep{kulk03}.  
Later {\it Chandra} and {\it XMM-Newton} data showed that the spin-down rate 
($\dot{P}$) might have decreased by $\sim$40\% over the course of several years 
\citep{tien09,guve12}. The estimated pulsed-fraction appeared to decrease to $f_p$ 
$\sim$ 4 \% by 2009 \citep{guve12}. Evidence of a slow decay ($\sim$20--30\% in 
$\sim$17 yr until 2009) in the soft X-ray flux was reported, which may be related 
to its long-term surface cooling and/or evolution of the X-ray emitting hot spot 
areas \citep{park12,guve12}.   
%$\dot{P}$ and X-ray flux 
%decays (as well as spectral softening) after the burst activities are expected 
%from untwisting magnetic fields in magnetar models (Thompson et al. 2002). 
%\citep[e.g.,][]{park12}

The 1 -- 10 keV band {\it XMM-Newton} spectrum can be fitted with a simple power 
law (PL) model with a steep photon index ($\Gamma$ $\sim$ 3.3) \citep{tien09}. 
The 0.5 -- 6 keV band {\it Chandra} spectrum (based on simultaneous spectral model
fits of six data sets taken in 2000, 2001, and 2009 with $\sim$30 -- 50 ks individual 
exposures) was fitted with spectral models for the X-ray emission from a magnetized 
H-atmosphere ($kT$ $\sim$ 0.35 keV) of a neutron star \citep{guve12}. The {\it Chandra} 
data with higher photon count statistics (with $\sim$110 ks effective exposure in total, 
combining four observations taken in 2009) showed that two-component models, e.g., 
thermal blackbody (BB) and/or nonthermal PL components, are required to adequately 
fit the observed X-ray spectrum in the 0.4 -- 8 keV band \citep{park12}. In the 
two-component spectral model fits, the soft component of the observed X-ray spectrum 
might be attributed to emission from the surface cooling neutron star (a BB-like 
emission with $kT$ $\sim$ 0.4 keV). Depending on the adopted spectral models, a hot 
spot(s) (e.g., BB with $kT$ $\sim$ 1 keV and $R$ $\sim$ 1 km) or the nonthermal 
magnetospheric radiation (e.g., a PL with $\Gamma$ $\sim$ 2.5) were required to fit 
the remaining part of the X-ray spectrum, primarily in the $\sim$3 -- 8 keV band. 

Broadband X-ray spectroscopy covering $E$ $>$ 10 keV is crucial to establish
the properties of the nonthermal component and to discriminate between the hotter 
thermal and nonthermal emission.
%to understand the 
%nature of the X-ray emission of 0526--66: i.e., a single vs. multiple spectral components, 
%and the origin of the harder spectral component and of the pulsed emission (thermal 
%vs. nonthermal). Identifying the emission source(s) is essential to study the neutron 
%star's cooling, magnetic field geometry, post-outburst evolution, and internal structure 
%processes. 
Here we report the results from our $\sim$50 ks {\it NuSTAR} observation, 
which provides the detection of the hard X-ray emission from 0526--66 up to $E$ $\sim$ 
40 keV. We also present results from our joint analysis of the broadband X-ray spectrum 
(in the 0.4 -- 40 keV band) of 0526--66 based on our {\it NuSTAR} and the archival {\it 
Chandra} data. We describe the observations and the data reduction in Section~2. 
We present our data analysis in Sections~3 \& 4, and a discussion in 
Section~5. A summary is presented in Section~6.

\section{\label{sec:obs} OBSERVATIONS \& DATA}

\begin{figure}[]
\figurenum{1}
\centerline{{\includegraphics[angle=0,width=\textwidth]{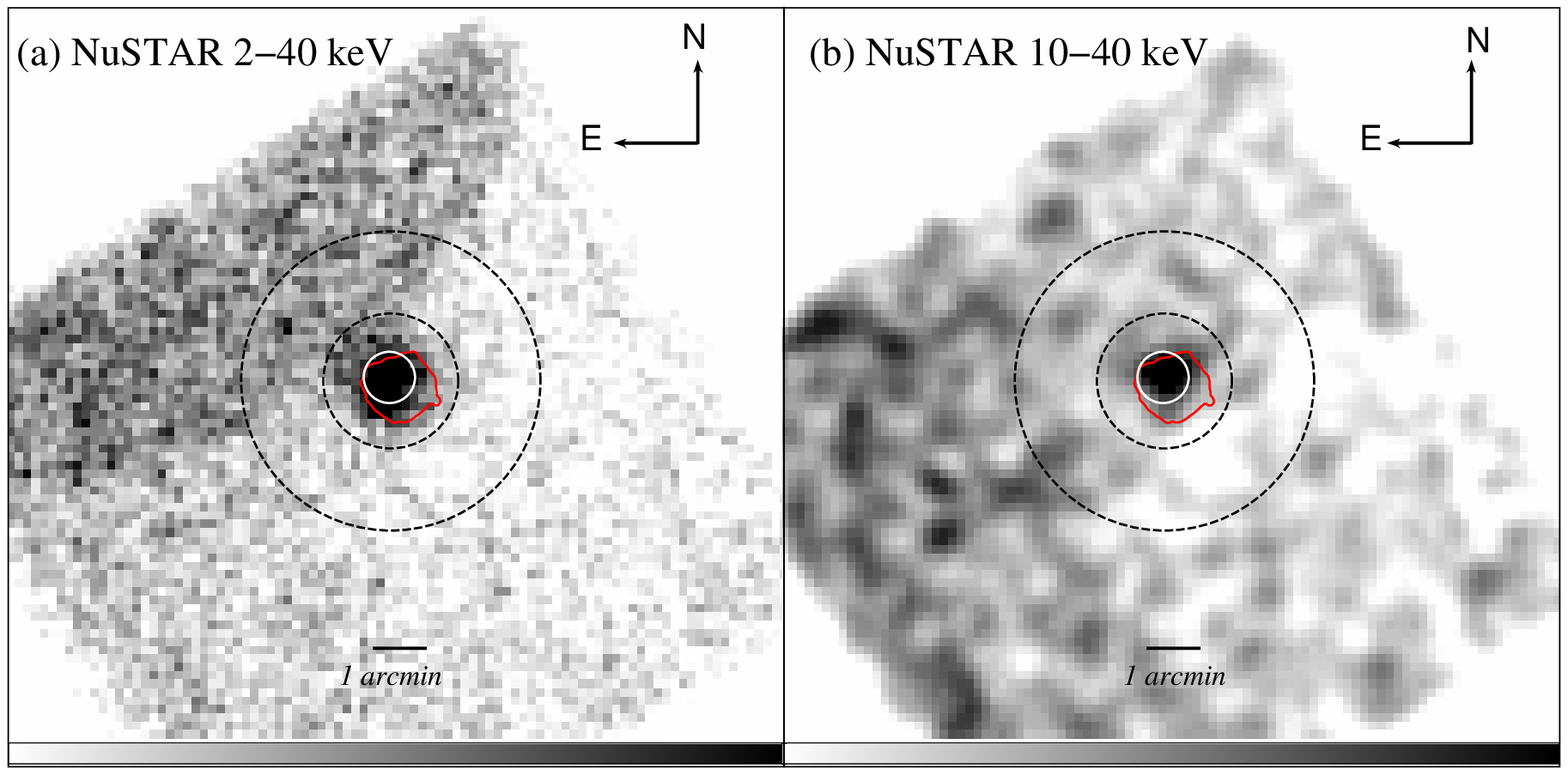}}}
%%{\includegraphics[angle=0,width=7.7cm]{fig1b.ps}}}
\figcaption[]{(a) Grey-scale {\it NuSTAR} images of SGR 0526--66: (a) the 2 -- 40 keV band,
and (b) the 10 -- 40 keV band. For the purposes of display both images have been binned by 
4 $\times$ 4 pixels. In (b), the image has also been smoothed. In (a) and (b), our source 
region for 0526--66 is marked with a white circle. The annular background region is marked 
with dashed circles. The outer boundary of the LMC SNR N49 (taken from the archival {\it
Chandra} ACIS data) is shown with red contours.   
\label{fig:fig1}}
\end{figure}

We observed 0526--66 with {\it NuSTAR} on 2018 February 7 during Cycle 3. We processed 
the data with NuSTARDAS (NuSTAR Data Analysis Software) version 1.7.1 and {\it NuSTAR}
CALDB version 20180126. After the standard data reduction, the effective exposure is 47 ks. 
We extracted the {\it NuSTAR} spectrum of 0526--66 from a circular source region with a radius
of 30$^{\prime\prime}$, centered on the source position RA (J2000) = 05$^h$ 26$^m$ 01$^s$.61, 
and Dec (J2000) = --66$^{\circ}$ 04$’$ 46$\farcs$0. We extracted the background spectrum from
an annular region around the source. The source and background spectra and spectral response 
files were created with the standard HEASARC software NUPIPELINE tool. Combining the data taken 
with two Focal Plane Modules (FPMA and FPMB), we extracted $\sim$1200 counts from the source
extraction region in the 2 -- 40 keV band. We estimate that $\sim$25\% of them are due to
the background, while the remaining ($\sim$900 counts) comprises both the SGR and emission 
from SNR N49 (Figure~\ref{fig:fig1} and Section~3). Based on our spectral model 
fits (see Section~3), we estimate 
the total background including the contamination by N49 is $\sim$60\% of the total extracted
counts in the 2 -- 40 keV band. In the 10 -- 40 keV band, we extract $\sim$280 
counts, $\sim$50\% of which is the background. The contamination from SNR N49 is negligible 
in this band. Thus, we make a clear $\sim$8$\sigma$ ($\sim$140 background-subtracted counts) 
detection of 0526--66 in the 10 -- 40 keV band.

About 1/4 of the field of view in the north-northeast of the source was moderately 
affected by scattered X-rays from nearby bright X-ray sources (Figure~\ref{fig:fig1}). The 
low-mass X-ray binary LMC X-3, which is projected at $\sim$2$\fdg$4 northeast of 0526--66, 
is the only bright X-ray source within a few degrees in the north-northeast of 0526--66. 
Thus, although LMC X-3 was not in a high state during our {\it NuSTAR} observation of 
0526--66 (based on the archival MAXI data), the contaminating source appears to be LMC X-3. 
0526--66 is positioned close to (or just outside of) the boundary of the affected background 
region (Figure~\ref{fig:fig1}). Based on our background spectral extractions from several 
regions around the source, we estimated a $\sim$10\% contamination from these scattered 
X-rays on the 10 -- 40 keV band source flux, which would not significantly affect our 
analysis.  In the following sections we assume the average background spectrum, extracted 
from an annular region around the source (Figure~\ref{fig:fig1}).

As supplementary data to model the broadband X-ray spectrum of 0526--66, we jointly analyzed 
the archival {\it Chandra} data of 0526--66 taken with the Advanced CCD Imaging Spectrometer 
(ACIS) in 2009 (ObsIDs 10123, 10806, 10807, and 10808)\footnote{This is a ``single'' observation,
which was split into four sub-sequences over a relatively short two-month period to accommodate 
the restrictions on the solar pitch angles for the telescope. There are several other {\it 
Chandra} archival data sets of 0526--66, which were taken several years earlier with shorter 
exposures than these 2009 data. Taking advantage of the longest effective exposure and the 
closest epoch of the observation (to that of our {\it NuSTAR} observation), we use these 2009 
{\it Chandra} data for the joint spectral analysis in this work. The large PSF of {\it 
XMM-Newton} resulted in a significant contamination from SNR N49 in the soft X-ray band 
spectrum of 0526--66 \citep{tien09}. Thus, we do not use the archival {\it XMM-Newton} data 
in this work.}. A 1/4 subarray of the ACIS-S3 detector was used in these 
archival {\it Chandra} data, which ensured a low photon pileup ($<$ 5\%) for a reliable spectral 
analysis of 0526--66. We processed each of the raw data sets and merged them, 
following the standard data reduction methods as described in Park et al. (2012). We extracted 
the source spectrum from a circular region centered on the source position (2$^{\prime\prime}$ 
in radius). Since 0526--66 is projected within the boundary of the X-ray emitting shell of SNR 
N49, the background characterization based on the high resolution imaging spectroscopy with 
{\it Chandra} ACIS data is critical. We extracted the background spectrum from an annular 
region around the source within SNR N49. Based on these data we extract 
$\sim$21000 photons (including $\sim$12\% background) for 0526--66 
in the 0.4--8 keV band.

\section{\label{sec:result} X-Ray Spectral Analysis}

The main goal of the {\it NuSTAR} observation of 0526--66 is to search for hard X-ray emission 
at $E$ $\ga$ 10 keV. While we make a clear detection of 0526--66 in the 10 -- 40 keV band, 
the utility of the {\it NuSTAR} data alone to adequately characterize the soft X-ray spectrum 
of 0526--66 is limited, because of the poor count statistics in the 
soft X-ray band (i.e., no response at $E$ $<$ 2 keV for FPM and the significant contamination 
from SNR N49's thermal emission at $E$ $<$ 5 keV). Thus, we perform 
the {\it NuSTAR} + {\it Chandra} joint spectral analysis to fit the broadband X-ray spectrum 
of 0526--66. For our spectral analysis we rebinned both of {\it NuSTAR} 
and {\it Chandra} spectra to contain a minimum of 20 counts per energy channel, and perform
all spectral model fits in the 0.4 -- 40 keV band. In our spectral model fits, we fixed the 
Galactic column at $N_{\rm H,Gal}$ = 6 
$\times$ 10$^{20}$ cm$^{-2}$ toward 0526--66 \citep{dick90}. We fit the foreground column in 
the LMC ($N_{\rm H,LMC}$) assuming the LMC abundances available in the literature \citep[He 
= 0.89, C = 0.30, N = 0.12, O = 0.26, Ne = 0.33, Na = 0.30, Mg = 0.32, Al = 0.30, Si = 0.30, 
S = 0.31, Cl = 0.31, Ar = 0.54, Ca = 0.34, Cr = 0.61, Fe = 0.36, Co = 0.30, and Ni = 0.62, 
][]{russ92,hugh98}. Hereafter, elemental abundances are with respect to solar \citep{ande89}. 
We tested recent X-ray measurements of the LMC abundances for O, Ne, Mg, and Fe, which were 
based on the X-ray spectral model fits of the shocked interstellar medium in the LMC SNRs 
\citep{magg16,sche16} for our $N_{\rm H,LMC}$ parameter. These LMC abundance measurements 
are lower than the previous values by $\sim$30--50\%, which resulted in slight increases 
(by $\sim$10--20\%) for the best-fit $N_{\rm H,LMC}$. Otherwise, the impact of these LMC 
abundances on the results from our spectral model fits of 0526--66 is not statistically 
significant, and thus does not affect our conclusions. For self-consistent comparisons 
with the previous results in the literature, we assumed the LMC abundances listed above 
in this work. 

In our spectral model fits, we tied $N_{\rm H, LMC}$, the BB temperature, the BB-emitting 
area, the PL photon index and normalization (see Sections~\ref{subsec:therm} \& 
\ref{subsec:nontherm} for our adopted spectral models) between the {\it NuSTAR} and {\it 
Chandra} spectra. At $E$ $\la$ 7 keV, the contamination from the soft thermal X-ray emission 
from SNR N49 (whose angular extent is $\sim$1$'$, Figure~\ref{fig:fig1}) is substantial in 
the {\it NuSTAR} spectrum of 0526--66 due to the large PSF (58$^{\prime\prime}$ HPD, Harrison 
et al. 2013) of {\it NuSTAR}. To account for this SNR spectrum, we added a plane-shock (PS) 
component (with the electron temperature of the hot gas $kT$ $\sim$ 1 keV on average, based 
on Park et al. [2003] and Uchida et al. [2015]) in the spectral modeling of the {\it NuSTAR} 
spectrum of 0526--66. We find that this background emission component from SNR N49 contributes 
$\sim$40\% of the {\it NuSTAR} flux of 0526--66 in the 2 -- 7 keV band, and is negligible 
($<$5\% of the total flux) in the 7 -- 40 keV band. Thanks to the sub-arcsecond resolution 
of the {\it Chandra} ACIS data, the contamination from the SNR in the {\it Chandra} spectrum
of 0526--66 is small. We subtracted it from the total spectrum of the source 
(Section~2), rather than applying this additional PS model component for the 
{\it Chandra} spectrum.

\subsection{\label{subsec:therm} Thermal Spectral Models for Hard Component}

There are several publicly-available spectral models for the X-ray spectrum emitted
by the neutron star's hot atmosphere with various atomic compositions, such as the NSA
model \citep{pavl95} and  NSMAX models \citep{ho08,mori07}. While these atmospheric models 
are more physically-motivated to describe the observed X-ray spectra from neutron 
stars, they are not adequate for magnetars, whose magnetic fields are very strong: 
e.g., these models allow the magnetic fields up to $B$ $\sim$ 10$^{13}$ G, whereas 
$B$ $\sim$ 10$^{14-15}$ has been estimated for 0526--66. Thus, we chose to use simple 
BB models for the thermal component of the X-ray spectrum of 0526--66 in this work,
realizing that they may provide only a phenomenological (rather than physically accurate)
description of the data. We note that fits with the BB model still allow us to investigate 
the thermal vs. nonthermal nature of the X-ray spectrum, and would also enable
self-consistent comparisons with the results from the magnetar analysis in the literature 
(where the BB models are commonly adopted). 

\begin{figure}[]
\figurenum{2}
\centerline{\includegraphics[angle=0,width=\textwidth]{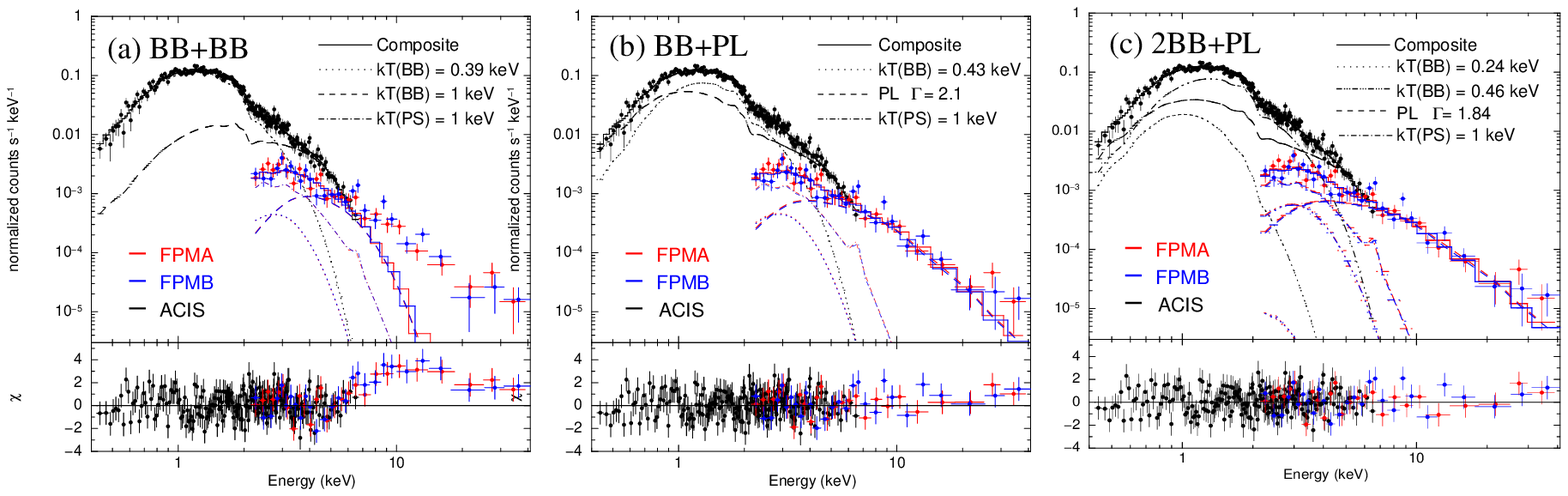}}
\figcaption[]{The observed {\it NuSTAR} and {\it Chandra} spectra of 0526--66.
The best-fit models are overlaid in each panel: (a) 2BB, (b) BB+PL, and (c) 2BB+PL.  
The PS component (with $kT$ = 1 keV, representing the contamination by thermal 
X-ray emission from SNR N49) is shown for the {\it NuSTAR} spectrum. The lower 
panels are residuals from the best-fit model. 
\label{fig:fig2}}
\end{figure}

In agreement with the previously published spectral analysis of the same {\it Chandra}
observations \citep{park12}, we find that the single BB model cannot fit the observed 
broadband X-ray spectrum of 0526--66 ($\chi^2$/$\nu$ $\sim$ 3). 
Then, we applied two component spectral models. %\citep[e.g.,][]{park12}. 
To test the thermal origin of the hard component of its X-ray spectrum, we fitted 
the broadband X-ray spectrum of 0526--66 with a two-component BB model. In this 
spectral model fit, we initially fixed the cool and hot component BB temperatures 
at the values derived by Park et al. (2012): i.e., $kT_{soft}$ $\approx$ 0.4 keV 
and $kT_{hard}$ $\approx$ 1 keV. The best-fit model is statistically poor 
($\chi^2$/$\nu$ $\sim$ 1.5) because our {\it NuSTAR} spectrum at $E$ $\ga$ 7 keV 
cannot be fitted with the BB model. When it is fitted, the temperature of the hot 
BB component increases to $kT$ $\approx$ 2 keV for the improved best-fit ($\chi^2$/$\nu$ 
$\sim$ 1.3). However, the systematic excess in the {\it NuSTAR} spectrum is still evident 
at $E$ $\ga$ 10 keV ($\chi^2$/$\nu$ = 2.7 in the 10 -- 40 keV band), even with this high 
temperature (Figure~\ref{fig:fig2}a). Thus, we conclude that the two-temperature BB model 
cannot adequately describe the {\it NuSTAR} + {\it Chandra} spectrum of 0526--66.

\subsection{\label{subsec:nontherm} Nonthermal Spectral Models for Hard Component}

\begin{figure}[]
\figurenum{3}
\centerline{\includegraphics[angle=0,width=\textwidth]{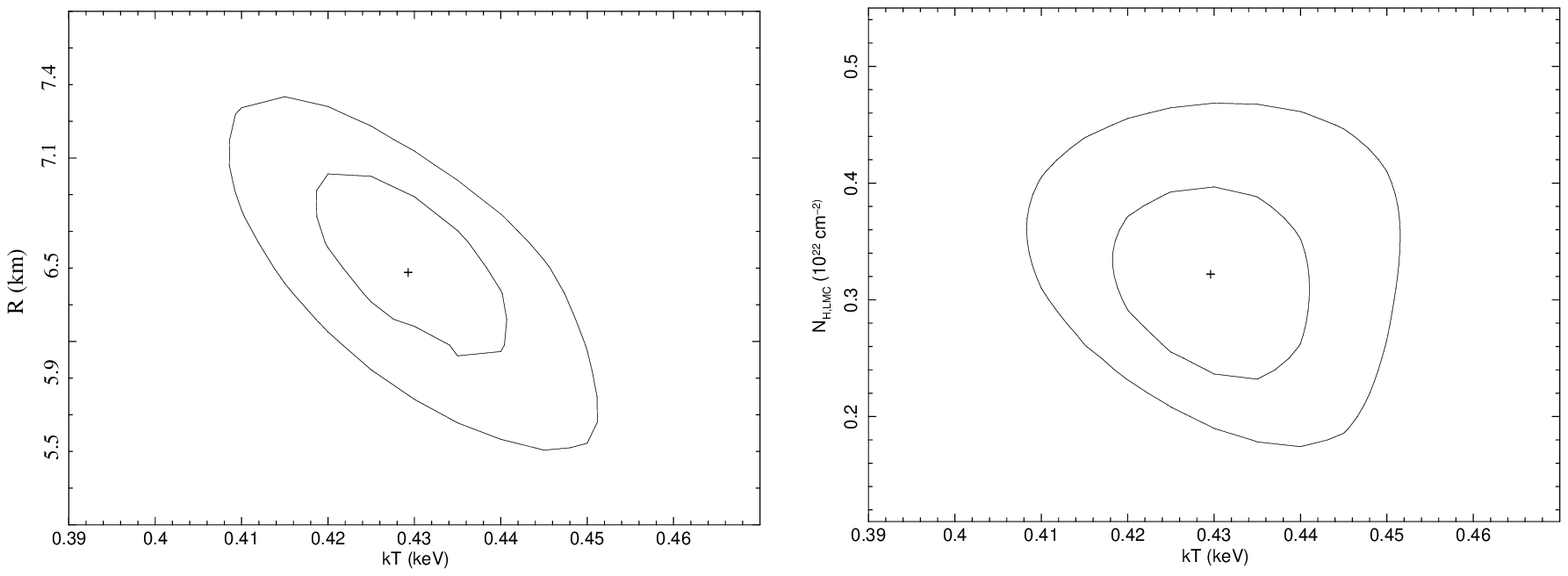}}
\figcaption[]{(a) The BB temperature vs. emission region radius (at $d$ = 50 kpc), 
and (b) the BB temperature vs. $N_{\rm H,LMC}$ contour plots based on our best-fit 
BB + PL model (Table~\ref{tbl:tab1}). In (a) and (b), 90\% and 99\% 
contours are shown. The best-fit temperature and emitting radius
are marked with a cross. 
\label{fig:fig3}}
\end{figure}

%To test the spectral nature of 0526--66 involving non-thermal components as suggested in 
%the previous {\it Chandra} and {\it XMM-Newton} data analysis \citep{kulk03,tien09,park12}, 
We first attempted a single PL model fit to the broadband X-ray spectrum of 0526--66.
The best-fit PL model is statistically unacceptable ($\chi^2$/$\nu$ = 1.8), showing
significant residuals both in the soft ($E$ $<$ 2 keV) and the hard bands ($E$ $>$ 5 keV). 
Next, we fitted the broadband X-ray spectrum of 0526 with BB + PL models. 
%Following the
%previous results \citep{park12}, we initially fitted the hard component spectrum of 
%0526--66 with a PL with the photon index $\Gamma$ = 2.5, while fixing the soft component 
%BB temperature at $kT_{BB}$ = 0.44 keV. This resulted in a statistically acceptable 
%($\chi^2$/$\nu$ = 1.13). When we varied the BB temperature and the PL photon index, $\Gamma$, 
The best-fit model results in $kT_{BB}$ = 0.43 keV and the PL photon index $\Gamma$ = 2.10 
($\chi^2$/$\nu$ = 1.09, Figures~\ref{fig:fig2}b \& \ref{fig:fig3}). 
%While the statistical improvement in 
%the model fit is marginal (F-probability $\sim$ 0.01), we note that the overall statistics 
%are dominated by the {\it Chandra} spectrum (with an order of magnitude higher count 
%statistics than those of the {\it NuSTAR} data). Also, since our main goals include 
%characterizing the hardband X-ray spectrum of 0526--66 for the first time, we take this 
%spectral model fit (with $kT_{BB}$ and $\Gamma$ varied) as the best-fit BB + PL model 
%in this work. Following a similar approach, we find that the CBB + PL model fit is 
%statistically acceptable ($\chi^2$/$\nu$ = 1.10), adequately describing the broadband 
%X-ray spectrum of 0526--66 (Figure~\ref{fig:fig2}c). In this CBB + PL model fits, the 
%best-fit BB temperature and PL photon index are $kT_{BB}$ = 0.43 keV and $\Gamma$ = 2.13, 
%respectively. 
This best-fit BB + PL model indicates an emission radius of the neutron star $R_{\rm NS}$ 
$\approx$ 6.5 km. These results are summarized in Table~\ref{tbl:tab1}.    

%Following the same approach as in \S~\ref{subsec:therm}, we then applied the NSA model, 
%instead of a BB, for the soft thermal component. In this NSA + PL model fit, we fixed 
%the PL photon index at $\Gamma$ = 2.5,  and assumed $M_{NS}$ = 1.4 $M_{\odot}$, $R_{NS}$ 
%= 10 km, $B$ = 10$^{13}$ G, and $d$ = 50 kpc for the neutron star. The best-fit model is 
%statistically acceptable ($\chi^2$/$\nu$ $\sim$ 1.13) with $kT_{NSA}$ = 0.39 keV. When 
%$\Gamma$ was varied, the best-fit value is $\Gamma$ = 2.4 with nearly the same $\chi^2$/$\nu$ 
%$\sim$ 1.13. Our NSMAX + PL model fits are also statistically acceptable ($\chi^2$/$\nu$ = 
%1.15) with similar values for the best-fit parameters.  The best-fit NSMAX model for the 
%H-rich atmosphere indicates the surface temperature of $kT$ $\sim$ 0.47 keV assuming $R$ = 
%10 km.  The photon index for the PL component is fixed at $\Gamma$ = 2.5. Varying $\Gamma$ 
%in the PL component results in a somewhat smaller best-fit value ($\Gamma$ $\sim$ 2.1) with 
%a $\sim$30\% lower $N_{H,LMC}$. The statistical improvement is, however, negligible with 
%the same reduced $\chi^2$. Our NSMAX model fits assuming the neutron star's atmosphere 
%enriched in intermediate-Z elements (C-, O-, or Ne-rich atmosphere [Mori \& Ho 2007]) are 
%statistically unacceptable ($\chi^2$/$\nu$ $\sim$ 1.8). 

Recent {\it NuSTAR} observations of magnetars and anomalous X-ray pulsars have shown 
the presence of similar hard X-ray emission components at $E$ $>$ 10 keV, for which
the broadband X-ray spectrum was fitted with three-component (2BB + PL) spectral models 
\citep[e.g.,][]{gott19}. Taking a similar approach, we apply 2BB + PL model fits for
0526--66. In these model fits, we assume the canonical radius of neutron stars ($R_{\rm NS}$ 
= 10 km) for the emitting area of the soft component BB spectrum. The best-fit model 
($\chi^2$/$\nu$ = 1.09) implies $kT_{\rm TH1}$ = 0.24 keV for the cooling surface 
of the neutron star. The best-fit hot spot temperature is $kT_{\rm TH2}$ = 0.46 keV with 
$R$ $\approx$ 6 km for the emitting area. In this model fit, a slightly harder PL component 
with the best-fit $\Gamma$ = 1.84 is obtained. Although the statistical improvement in 
the spectral fit with the 2BB + PL model is marginal (the F-probability $\sim$ 0.02), 
the consideration of a cooler emission component from the neutron star with the canonical 
size may be useful to understand the physical nature of 0526--66, and thus we include the 
results from this 2BB + PL model fit in Table~\ref{tbl:tab1}.

\section{\label{sec:timing} {\it NuSTAR} Timing Analysis}

Due to poor photon statistics and the large background ($\sim$500 counts per FPM in 
the 2 -- 40 keV band with $\sim$60\% background [including the contamination from N49]), 
the pulsation search (for the presumed small $f_p$ $\la$ 10\%) is difficult with 
our {\it NuSTAR} data. Based on the previously reported values of $\dot{P}$ and $P$, 
we calculated $Z_1^2$ statistic as a function of frequency within a plausible range of  
$P$ = 8.044 -- 8.076 s \citep{guve12} using events from both FPMA and FPMB with the 
arrival times corrected for the Earth and spacecraft motion. We did not find any 
statistically significant signal with the  two largest values of $Z_{1,\rm max}^2$ 
$\approx10$ (both are found near $8.05348$ s). Therefore, we calculate an upper limit 
on the pulsed fraction (for a sinusoidal pulse profile with a single peak per period) as 
$f^{\rm obs}_{p,n\sigma}\approx2\left[\ln(\mathcal{N}/P_n)/N\right]^{1/2}$ where, $N=1159$ 
is the total number of events (source plus background), $\mathcal{N}=183$ is the number 
of independent frequency trials, and $P_n={\rm erfc}(n/\sqrt{2})\approx(2/\pi)^{1/2}e^{-n^2/2}/n$ 
(for $n>2$) with $n$ being the confidence level in units of standard deviation $\sigma$. 
For $n=3$, $P_n\approx0.003$ and, hence, $f^{\rm obs}_{p,3\sigma}=0.195$ is an upper 
limit on the observed pulsed fraction at $3\sigma$ confidence. The corresponding 
intrinsic pulsed fraction is $f^{\rm int}_{p,3\sigma}=f^{\rm obs}_{p,3\sigma}(N/N_{\rm 
bg})=0.48$ because the background is contributing $60\%$ of the total events within the 
source extraction aperture, $r$ = 30$^{\prime\prime}$, in the 2 -- 40 keV band.

\section{\label{sec:disc} DISCUSSION}

\subsection{\label{subsec:0526} Broadband X-Ray Spectral Nature of 0526--66}

Based on our {\it NuSTAR} and {\it Chandra} data analysis, we find that a PL-like 
nonthermal spectral component, in addition to the soft thermal BB component, is required 
to adequately fit the broadband X-ray spectrum of 0526--66. In our BB + PL model fits, 
a BB temperature of $kT$ $\approx$ 0.43 keV is estimated, where the implied radius 
of the emission region ($R$ $\approx$ 6.5 km at the distance of $d$ = 50 kpc) is smaller 
than the canonical neutron star size ($R$ $\sim$ 10 km). The best-fit 2BB + PL model shows
a lower BB temperature of $kT$ = 0.24 keV, assuming the neutron star radius of $R_{\rm NS}$ 
= 10 km. The cooler ($kT$ = 0.24 keV) component BB flux contributes $\sim$15\% of the total 
BB luminosity in the 0.5 -- 5 keV band. Thus, for both of the BB + PL and 2BB + PL model 
fits, the BB component with $kT$ $\sim$ 0.4 -- 0.5 keV dominates the thermal component 
X-ray emission of 0526--66. Spin-down ages of $\tau_{sd}$ $\sim$ 2000 -- 3400 yr have 
been estimated for 0526--66 \citep{kulk03,tien09,guve12}. While the physical association between
SGR 0526--66 and SNR N49 is in debate (see Section~\ref{subsec:nh}), these spin-down 
ages are in plausible agreement (within a factor of $\sim$2) with the estimated Sedov 
age of SNR N49 ($\tau_{Sedov}$ $\sim$ 4800 yr, Park et al.  2012). Thus, we may estimate
the age of 0526--66 to be $\tau$ $\sim$ 10$^{3-4}$ yr. If either of these BB temperatures 
corresponds to that of the passively cooling neutron star's surface after its birth, 
0526--66 is hotter than the predictions of the cooling curves for neutron stars of this
age \cite{yako04}. For a neutron star in this age range, the estimated high surface temperature 
may suggest a re-heating of the surface, probably due to the recent burst activity, and/or 
strong magnetic field effects on the thermal evolution of the neutron star \citep{pons09,viga13}.    

Based on our best-fit spectral models, we estimate that the hardness ratio for the
unabsorbed X-ray flux (HR = $f_{\rm 15-60~keV}$/$f_{\rm 1-10~keV}$ as utilized by 
Enoto et al. [2017]) is HR $\sim$ 0.3--0.4 (depending on the adopted model). The latest 
estimates for the time-derivative of the pulsation period, the magnetic field strength, 
and the characteristic age for 0526--66 are $\dot{P}$ $\sim$ 4 $\times$ 10$^{-11}$ s 
s$^{-1}$, $B$ $\sim$ 4 $\times$ 10$^{14}$ G, and $\tau_{sd}$ $\sim$ 3200 yr, respectively 
\citep{guve12}. We compare our estimated HR values for 0526--66 with the PL-like 
empirical relations with $\dot{P}$, $B$, and the neutron star's age, which were 
suggested based on the Galactic sample of a $\sim$dozen magnetars \citep{enot17}. 
Our estimated range of HR for 0526--66 is lower than HR $\sim$ 1 that is predicted 
by Enoto et al. (2017). $\dot{P}$ (and thus $B$) for 0526--66 appears to be decreasing 
between 2000 and 2009 \citep{kulk03,tien09,guve12}. The soft X-ray flux of 0526--66 has 
been decaying between 1992 and 2009 \citep{park12,guve12}. It is difficult to estimate 
changes in the BB temperature of 0526--66 between 2009 and 2018.  The {\it NuSTAR}'s 
detector response is limited to $E$ $>$ 2 keV, and the soft X-ray spectrum (at $E$ $\la$ 
5 keV) of 0526--66 in the {\it NuSTAR} data is significantly contaminated by that from 
SNR N49. Based on the BB + PL model fit to our {\it NuSTAR} data alone with an assumption 
that the BB emitting area stays the same as it was in 2009, we place a 2$\sigma$ upper 
limit of $kT_{BB}$ $\la$ 0.5 keV for the thermal component emission of 0526--66 as of 
2018. Considering the possible temporal evolution in $\dot{P}$, $B$, and $kT$, we 
speculate that it might be partially responsible for the low HR. 
Our estimated limits on $kT_{BB}$ and $f_p$ as of 2018 based on our 
{\it NuSTAR} data are not constraining 
due to poor photon statistics. Significantly deeper {\it NuSTAR} and {\it Chandra} 
observations would be required to constrain the recent evolution of these critical neutron 
star parameters.
%The corresponding intrinsic pulsed fraction 
%(for a sinusoidal pulse 
%profile with single peak per period can be estimated as 
%pf=sqrt(2 Z_1,max^2/Ntot)*(Ntot/Nsrc)=sqrt(2 Z_1,max^2/Ntot)*(1/(1-Nbg/Ntot)) = 31 %
%here (Ntot/Nsrc) is the correction from observed to  intrinsic pf,  Z_1,max^2=9,  
%Nttot=Nsrc+Nbg=1159 (from FPMA and FPMB together) and I took Nbg/Nsrc=0.6 

Our spectral fits indicate the photon index $\Gamma$ $\sim$ 2 for the PL component 
which dominates the hard X-ray spectrum of 0526--66 at $E$ $\ga$ 10 keV. This spectral 
slope of the PL component is larger than those measured for other Galactic magnetars 
($\Gamma$ $\la$ 1.5, e.g., Enoto et al. [2017]). Apparently, 0526--66 has not shown 
any activity since its giant bursts in 1979, being in the quiescent state (probably 
with a steady cooling) for the longest time period among the observed magnetars. 
We speculate that this distinctive evolutionary state of 0526--66 (compared with 
that for other magnetars with more frequent burst activity) in the last several 
decades might have led to a steeper nonthermal spectral component, and thus somewhat 
lower HR in the latest broadband X-ray spectrum. 
%While the origin of our estimated HR for 0526--66 is unclear, 
%significantly deeper and contemporary {\it Chandra} and {\it NuSTAR} observations of 
%0526--66 are required to adequately trace the changes in the surface temperature, 
%$P$, and $\dot{P}$ of 0526--66 since 2009.
The best-fit PL photon index for 0526--66 is similar to that of the synchrotron 
emission spectrum from the relativistically accelerated electrons in pulsar 
magnetospheres, or from pulsar wind nebulae (PWNe, although some PWNe show harder 
spectra with $\Gamma$ $\sim$ 1.0 -- 1.5). 
%The bright Crab-like pulsar PSR 0540--69.3 in the LMC shows the extended 
%($\sim$6$^{\prime\prime}$ in radius) synchrotron nebula, for which the X-ray spectrum 
%can be fitted with PL models with similar $\Gamma$ ($\sim$ 1.4 -- 2.5, Petre et al. 2007) 
%to those estimated for 0526--66. The estimated soft X-ray luminosity of 0526--66 is 
%$\sim$50 times lower than that of 0540--69.3. Scaling the soft X-ray luminosities 
%between 0526--66 and 0540--69.3 predicts a considerable ACIS count rate of $\ga$0.01 
%c s$^{-1}$ for the immediate surrounding regions (within a radius of $\sim$5$^{\prime\prime}$) 
%of the detected pointlike source 0526--66. 
However, the high-resolution {\it Chandra} ACIS images show no evidence for an extended 
nebulosity around the pointlike source 0526--66 \citep{kulk03,park12}. If we entertained 
the intriguing possibility of a PWN being developed for 0526--66, the lack of an observed 
extended nebula in the {\it Chandra} data would require an angular size of $\la$1$^{\prime\prime}$ 
that cannot be resolved by {\it Chandra}. The corresponding physical size is $\la$0.2 pc 
at the distance of the LMC ($d$ = 50 kpc), which would be similar to (or smaller than) 
those of the torus and jet associated with young pulsars found in the Galactic SNRs 
G292.0+1.8 \citep{park07} and 3C58 \citep{slan02}. 

\subsection{\label{subsec:nh} Notes on Foreground Absorption for 0526--66}

SGR 0526--66 is projected within the boundary of SNR N49 in the LMC \citep{clin82,roth94}. 
However, the physical association between the two has not been conclusive. The statistical 
chance probability for a coincidental alignment along the line of sight between N49 and 
0526--66 is not negligible \citep{gaen01}. If the massive progenitor of 0526--66 was born 
in a nearby massive stellar cluster, a significantly older age ($\sim$10$^5$ yr) than that 
of N49 ($\sim$4800 yr) is implied for 0526--66 \citep{klos04}. In our spectral model fits, 
the contribution from the PL component is considerable in the soft band X-ray flux (at $E$ 
$<$ 2 keV) as well as in the hard band ($E$ $>$ 5 keV). This results in an inferred 
foreground LMC column $N_{\rm H,LMC}$ $\sim$ 3 $\times$ 10$^{21}$ cm$^{-2}$ toward 
0526--66 (Table~\ref{tbl:tab1}), a factor of $\sim$2 larger than that estimated for 
SNR N49. This higher LMC column for 0526--66 might cast further doubt on its physical 
association with N49. When we fixed the LMC column at the value estimated for N49 
($N_{\rm H,LMC}$ $\sim$ 1.5 $\times$ 10$^{21}$ cm$^{-2}$, Park et al. 2012), the 
overall spectral model fits (for those models presented in Table~\ref{tbl:tab1}) 
became slightly poorer ($\chi^2/{\nu}$ $\sim$ 1.14). Although these fits may be 
formally distinguished from those summarized in Table~\ref{tbl:tab1} (e.g., 
$F$-probability $\sim$ 6 and 1 $\times$ 10$^{-4}$, for the BB + PL and 2BB + PL model 
fits, respectively), the overall fits may still be considered to be statistically 
acceptable. Also, due to the model-dependence of these $N_{\rm H,LMC}$ estimates (between 
0526--66 and N49) and the relatively large uncertainties, it may not be straightforward 
to firmly conclude that the X-ray spectrum of 0526--66 is more absorbed than that of N49. 
%Nonetheless, with the assumed low $N_{\rm H,LMC}$, a $\sim$20\% smaller $\Gamma$ 
%($\sim$1.4 -- 1.7) are obtained for the PL component. Due to the implied lower column 
%and harder PL spectral slopes, higher hardness ratios of the unabsorbed flux (HR $\sim$ 
%0.5 -- 0.7) are estimated. These moderate changes do not significantly affect our 
%discussion above. 
%The best-fit parameter values are consistent (within uncertainties) with to those listed 
%in Table~\ref{tbl:tab1}.

\section{\label{sec:sum} SUMMARY}

We have confidently detected SGR 0526--66 in the quiescent state at $E$ $>$ 10 keV
with {\it NuSTAR}. The joint spectral fits to the {\it NuSTAR} + {\it Chandra}
spectrum require at least two component model if BB and PL are used for the individual 
components. The hard X-ray emission at $E$ $>$ 10 keV, which is fitted with a PL spectrum,
is most likely nonthermal in origin. The slope of the PL component ($\Gamma$ 
$\sim$ 1.8 -- 2.1) is softer than those for other magnetars observed with {\it NuSTAR}. 
For our best-fit BB + PL model, we obtain a high BB temperature of $kT$ $\approx$ 0.43 
keV, and the estimated radius of the X-ray-emitting area ($R$ $\approx$ 6.5 km) is smaller 
than the canonical size of a neutron star. If we assumed the BB spectrum from the cooling 
neutron star with a standard radius of $R_{\rm NS}$ = 10 km in addition to the hot $kT$ 
$\sim$ 0.4 keV component, a lower BB temperature of $kT$ $\approx$ 0.24 keV is obtained 
for the passively cooling surface of the neutron star. The presence of the relatively 
soft PL component in the X-ray spectrum of 0526--66 may provide an intriguing opportunity 
to study the magneto-thermal evolution of a magnetar during the substantially long 
quiescent period of several decades after the strong outbursts. Based on the {\it NuSTAR} 
data, we place a 3$\sigma$ upper limit of $f_p$ $\sim$ 0.48 on the intrinsic pulsed 
fraction of 0526--66 in the 2 -- 40 keV band.

\acknowledgments

We thank the anonymous referee for her/his comments that helped improving this 
manuscript. This work has been supported in part by NASA {\it NuSTAR} grant 80NSSC17K0633
and the {\it Chandra} grant GO9-0072A to the University of Texas at Arlington. 
O. K. was supported in part by NASA through {\it Chandra} Award number TM8-19005B
issued by the {\it Chandra} X-Ray Center which is operated by the Smithsonian
Astrophysical Observatory for and on behalf of NASA under the contract NAS8-03060.

\clearpage

\begin{deluxetable}{lccc}
\footnotesize
\tablecaption{Summary of spectral model fits to SGR 0526--66
\label{tbl:tab1}}
\tablewidth{0pt}
%\tablehead{ \colhead{Parameter} & \colhead{BB+PL} & \colhead{NSA+PL} & \colhead{NSMAX+PL} &
%\colhead{CBB+PL} & \colhead{2BB+PL}}
\tablehead{ \colhead{Model Parameter} & \colhead{BB + PL} & \colhead{2BB + PL}}
\startdata
%$kT_{TH1}$ (keV) & 0.44 (fixed) & 0.39$\pm$0.01 & 0.47$\pm$0.01 & 0.42$^{+0.04}_{-0.02}$ & 0.24$^{+0.05}_{-0.06}$ \\
$kT_{\rm TH1}$ (keV) & 0.43$\pm$0.01  & 0.24$^{+0.05}_{-0.06}$ \\
%$kT_{TH2}$ (keV) & -  & - & - & - & 0.46$^{+0.05}_{-0.03}$ \\
$kT_{\rm TH2}$ (keV) & - & 0.46$^{+0.05}_{-0.03}$ \\
%$\Gamma$ & 2.5 (fixed) & 2.5 (fixed) & 2.5 (fixed) & 2.5 (fixed) & 1.85$\pm$0.31 \\
$\Gamma$ & 2.10$^{+0.16}_{-0.22}$ & 1.84$\pm$0.24 \\
%$R_{\rm TH1}$ (km) & 5.5$\pm$1.6 & 10 (fixed) & 10 (fixed) & 7.7$^{+4.6}_{-4.1}$ & 10 (fixed) \\
$R_{\rm TH1}$ (km) & 6.5$\pm$0.5 & 10 (fixed) \\
%$R_{\rm TH2}$ (km) & - & - & - & - & 5.9$^{+0.8}_{-1.3}$ \\
$R_{\rm TH2}$ (km) & - & 5.9$^{+0.8}_{-1.3}$ \\
%$B_{\rm NS}$ (10$^{13}$ G) & - & 1 (fixed) & 3 (fixed) & - & - \\
%$f_{\rm TH}$\tablenotemark{a} & 3.4$\pm$0.3 & 3.6$\pm$0.1 & 4.4$\pm$0.1 & 4.9$^{+0.6}_{-1.1}$ & 5.6$^{+0.8}_{-1.3}$ \\
$f_{\rm TH}$\tablenotemark{a} (0.5 -- 5 keV, 10$^{-13}$ erg cm$^{-2}$ s$^{-1}$) & 4.5$^{+0.5}_{-0.8}$ & 5.7$^{+0.7}_{-1.2}$ \\
%$f_{\rm PL}$\tablenotemark{b} & 3.8$\pm$0.3 & 3.8$\pm$0.3 & 3.3$\pm$0.4 & 3.2$\pm$0.7 & 5.2$^{+1.0}_{-2.5}$ \\
$f_{\rm PL}$\tablenotemark{a} (5 -- 40 keV, 10$^{-13}$ erg cm$^{-2}$ s$^{-1}$) & 4.4$^{+0.5}_{-1.4}$ & 5.3$^{+0.8}_{-4.6}$ \\
%$L_{\rm TH}$\tablenotemark{a} & 1.4$\pm$0.1 & 1.5$\pm$0.1 & 1.9$\pm$0.1 & 1.9$^{+0.2}_{-0.4}$ & 2.2$^{+0.3}_{-0.5}$ \\
$L_{\rm TH}$\tablenotemark{b} (0.5 -- 5 keV, 10$^{35}$ erg s$^{-1}$) & 1.8$^{+0.2}_{-0.3}$ & 2.2$^{+0.3}_{-0.5}$ \\
%$L_{\rm PL}$\tablenotemark{b} & 1.1$\pm$0.1 & 1.1$\pm$0.1 & 1.0$\pm$0.1 & 1.0$\pm$0.2 & 1.6$^{+0.3}_{-0.7}$ \\
$L_{\rm PL}$\tablenotemark{b} (5 -- 40 keV, 10$^{35}$ erg s$^{-1}$) & 1.3$^{+0.1}_{-0.4}$ & 1.6$^{+0.2}_{-1.4}$ \\
%$f_{\rm broad}$\tablenotemark{c} (10$^{-13}$ erg cm$^{-2}$ s$^{-1}$) & 12.2$\pm$1.4 &
%12.2$\pm$1.2 & 12.7$\pm$1.5 \\
%$L_{\rm broad}$\tablenotemark{c} (10$^{35}$ erg s$^{-1}$) & 5.3$\pm$0.6 & 5.3$\pm$0.5
%& 5.4$\pm$0.6 \\
%$f_X$\tablenotemark{c} & 12.6$\pm$3.1 & 12.6$\pm$2.2 & 13.7$\pm$2.7 & 12.8$\pm$3.6 & 15.3$\pm$1.7 \\
$f_{\rm X}$\tablenotemark{a} (0.5 -- 60 keV, 10$^{-13}$ erg cm$^{-2}$ s$^{-1}$) & 14.1$^{+2.1}_{-4.2}$ & 15.4$^{+1.7}_{-5.9}$ \\
%$L_X$\tablenotemark{c}  & 5.4$\pm$1.3 & 5.4$\pm$1.0 & 5.6$\pm$1.1 & 5.1$\pm$1.5 &5.4$\pm$0.6 \\
$L_{\rm X}$\tablenotemark{b} (0.5 -- 60 keV, 10$^{35}$ erg s$^{-1}$) & 5.2$^{+1.2}_{-1.6}$ & 5.4$^{+1.0}_{-3.4}$ \\
%$N_{\rm H,LMC}$ (10$^{21}$ cm$^{-2}$) & 5.1$\pm$0.3 & 5.1$\pm$0.3 & 4.9$\pm$0.2 & 4.1$^{+0.5}_{-1.1}$ & 2.8$^{+0.9}_{-0.7}$ \\
$N_{\rm H,LMC}$ (10$^{21}$ cm$^{-2}$) & 3.4$\pm$1.0 & 2.7$^{+0.7}_{-0.6}$ \\
%$N_{\rm H,Gal}$ (10$^{21}$ cm$^{-2}$) & - & - & 0.6 (fixed) \\
%$\chi^2$/$\nu$ & 316.3/280 & 317.1/280 & 321.6/279 & 313.0/278 & 301.0/277
$\chi^2$/$\nu$ & 305.5/279 & 299.8/278
\enddata
\tablecomments{Uncertainties are with a 90\% C.L. The Galactic column is fixed at 
$N_{\rm H,Gal}$ = 6 $\times$ 10$^{20}$ cm$^{-2}$. $d$ = 50 kpc is assumed. }
\tablenotetext{a}{The observed flux.} 
\tablenotetext{b}{The X-ray luminosity after removing the absorption.} 
\end{deluxetable}

\end{document}